\documentclass{sig-alternate}
\usepackage{graphicx}
\graphicspath{{figs/}}
\usepackage{subfigure}

\begin{document}

\date{}

\title{Bridge the Gap: Measuring and Analyzing Technical Data for Social
Trust between Smartphones}

\numberofauthors{3}
\author{
\alignauthor Sebastian Trapp\\
  \affaddr{Institut f\"ur Informatik\\Freie Universit\"at Berlin\\14195 Berlin, Germany\\
 sebastian.trapp@fu-berlin.de}
\alignauthor Matthias W{\"a}hlisch\\
  \affaddr{Institut f\"ur Informatik\\Freie Universit\"at Berlin\\14195 Berlin, Germany\\
 m.waehlisch@fu-berlin.de}
\alignauthor Jochen Schiller\\
  \affaddr{Institut f\"ur Informatik\\Freie Universit\"at Berlin\\14195 Berlin, Germany\\
  jochen.schiller@fu-berlin.de}
}

\conferenceinfo{} {}
\CopyrightYear{}
\crdata{}

\maketitle

\begin{abstract}

Mobiles are nowadays the most relevant communication devices in terms of quantity
and flexibility. Like in most MANETs ad-hoc communication between two mobile phones requires mutual trust between the devices.
A new way of establishing this trust conducts social trust from technically measurable data (e.g., interaction logs).
To explore the
relation between social and technical trust, we conduct a large-scale
survey with more than 217 Android users and analyze their anonymized call
and message logs. We show that a reliable a priori trust value for a mobile
system can be derived from common social communication metrics. 

\end{abstract}



\section{Introduction}

The ever increasing distribution of smartphones creates new networking
possibilities. Equipped with several network interfaces the mobiles can form
spontaneous ad hoc networks (e.g., to bypass bad links). Connecting to
another device as sender, receiver, or relay for data packets, always requires
trust towards the peer, as reciprocal dependencies arise concerning
data privacy, service availability, and incentives for investing energy.
Determining the trustworthiness of nearby devices is an important but also
challenging requirement for \emph{ad hoc} communication as central
instances or pre-shared secrets are not feasible.

New security approaches \cite{abkw-vsff-08,tws-spypt-11} arise that try to
apply social trust of the phones' users to derive technical trust between
devices. Bridging the gap between such social trust and measurable data on
the devices is still an open research topic. On the one hand, complete
logs and directories stored at the mobile are required to explore the data
space. On the other hand, a detailed understanding of appropriate
metrics for trust calculation is bound to the direct interaction with the
end user (i.e., questionnaires) at least for verification.

In this paper, we conduct a large-scale survey and measurement with over
217~Android users to identify common communication parameters that are
available at mobiles and that can indicate trust relationships between
individuals. The contributions of this paper are the following: (1) We
analyze more than 22,365~calls and 84,325~messages of a globally
distributed set of subjects with respect to the general communication
behaviour and trust establishment; (2) we map social trust on technical
trust; (3) we evaluate trust prediction between smartphones. From a
network perspective, we use the (maybe unusual) methodology of a survey.
However, we explicitly note that the ranking of communication patterns
according to trust requires a feedback from the users.

Current studies either use a provider view on mobile phone communication
but do not provide user specific data comprising hints about inter-personal
relationships. User-centric studies can provide those insights but
include only very limited measurements at the moment.

The remainder of the paper is structured as follows: In Section
\ref{sec:related-work} we describe and analyze the problem space and
discuss other studies using mobile phone contact and communication data. 
The design and implementation of our survey are presented in Section \ref{sec:survey} with a subsequent analysis of its results in section \ref{sec:evaluation}.
Section \ref{sec:application} briefly describes application perspectives for the results obtained in this study. 
We conclude with an outlook in Section \ref{sec:conclusion}.

\section{Problem Statement \& Related \\ Work}\label{sec:related-work}

With his seminal work, Granovetter~\cite{g-swt-73} established \emph{tie strength}
as a measure for the relationships between individuals in their social networks. On this basis he showed the transitivity of trust, i.e., if two individuals A and B are strong ties (e.g., good friends, family, etc.) with person C, there also exists a tie between A and B directly.
Transitive trust provides the basis for a new class of communication protocols for mobile phones. 
Especially in the context of phones forming mobile ad-hoc networks
(MANETs),
ad-hoc trust establishment is needed. When comparing contact list data
trust can transitively be assigned in a privacy-friendly way \cite{tws-spypt-11}. Friends of friends
detection in mobile networks also relies on transitive trust \cite{abkw-vsff-08}.
Those protocols work with the contact list data or communication logs available on every mobile phone.
In order to exploit this feature the tie strength towards other individuals has to be determined. Closeness or emotional intensity of a relationship are good
indicators for strong ties \cite{mc-mts-84} but they are not directly defined in observable data. On the other hand, duration and frequency of communication describe explicitly social interaction and can be discovered from call logs. Closeness, duration, and frequency all are positively related to an \emph{intimacy} component of communication \cite{m-ndm-90}, combining all aspects increases the confidence in strong ties.
While many of the previous studies were performed before the era of widely distributed mobile phones recent data is needed to create a mapping between technical measurable data (e.g., call and message logs) and the associated tie strength or trust attribute.

Current studies that could be of use to associate technical data of mobile
phones with social trust of the phones' users are either provider centric
\cite{smsbf-mcgbp-08, oshsl-stsmc-07} or surveyed locally on the mobile
phones themselves \cite{ep-rmscs-06, pd-tucwc-11}. Operator-based
measurements are based on extensive data but rely on access to centrally
observed information, which cannot be assumed in a deployed ad hoc scenario. Furthermore, they either provide
only call records of intra-provider communication \cite{oshsl-stsmc-07} or
those collected at designated base station controllers
\cite{smsbf-mcgbp-08} containing only the communication activities that
took place in geographic vicinity to those switches. In addition, data
collected at the provider level lacks locally available meta data such as
tags assigned by the user to contact entries (e.g., favorites, home or
work numbers). These tags can provide valuable information in tie strength
analysis. Local, mobile centric data collection can provide such a rich depth
of information and generally comprises a complete set of communication data
of the specific user, not restricted by measuring points. 
The current data sets from mobile phones, however, contain only data of at
most 100 participants and analyze only a limited set of metrics \cite{ep-rmscs-06}.

To gain the information needed for protocols using the transitivity of
trust, all call and message logs available on a mobile phone should be
comprised. In order to come to a conclusion regarding tie strength or trust
values towards peers, the participants of the study should be able to individually rate
their contacts.

\section{Exploring Social Trust at Mobiles}\label{sec:survey}

\newcommand{\numResultsAMT}{257 }
\newcommand{\numResultsAMTUnique}{220 }
\newcommand{\numResultsAMTHonest}{217 }

\newcommand{\numPartnersAMT}{7,114 }
\newcommand{\numCallsAMT}{22,365 } 
\newcommand{\numMsgsAMT}{84,325 } 
\newcommand{\numPartnersRatedAMT}{3,449 }
\newcommand{\numResultsAMTClean}{188 }
\newcommand{\numCallsAMTClean}{22,157 } 
\newcommand{\numMsgsAMTClean}{83,941 } 
\newcommand{\numPartnersAMTClean}{6,992 }
\newcommand{\numPartnersRatedAMTClean}{3,331 }


We conducted a quantitative survey to explore the relationship between
measurable features on mobile phones and closeness or trust assigned to
corresponding communication partners. Analyzing the answers of the subjects
in combination with gathered local data at the mobiles allows us to predict
closeness and trust. Our empirical study further examines how closeness and
trust correlate in a mobile phone's context.

\subsection{Basic Survey Setup} \label{ssec:surveySetup}

To overcome the problems that arise from pre-defined data or biasing
environments, we implemented the survey as an application for smartphones.
The app has direct access to the local communication data and thus allows
to study to which extent data is available on real mobiles and how they
describe trust and closeness. For selected entries in their contact list,
participants were asked to respond to each of the following statements:

\begin{description}
  \item[Closeness:] \emph{I feel close to this person.}
  \item[Trust information:] \emph{I would trust this person with sensitive information.}
  \item[Trust best:] \emph{I trust that this person wants the best for me.}
\end{description}

The first item evaluates closeness, a common measure in relationship
analysis (cf., Section~\ref{sec:related-work}). The two subsequent
statements aim to evaluate two characteristics of trust: The trust
regarding sensitive information is to be given, a very important property
in the context of ad hoc networks, and from a more general perspective the
overall trust between two people. The responses range from \emph{strongly
disagree} (rating 1) to \emph{strongly agree} (rating 5) on a five point
Likert~scale~\cite{l-tma-32}. Combining all three statements reveals if
\emph{closeness} to an individual is related to \emph{trust}.

Conducting a large-scale survey is non-trivial, in particular when
sensitive information such as real contact data will be used. To
successfully recruit participants, the survey (app) should preserve privacy
and operate transparently, as well as give motivation to participate.

\subsubsection{Survey Application} \label{ssec:application}

We developed the survey app for Google Android because it allows the
installation of applications outside the Android Market and reaches a
variety of people due to the wide range of models and device prices.

\begin{figure}
  \center
  \subfigure[Rating screen]{\includegraphics[width=0.39\columnwidth]{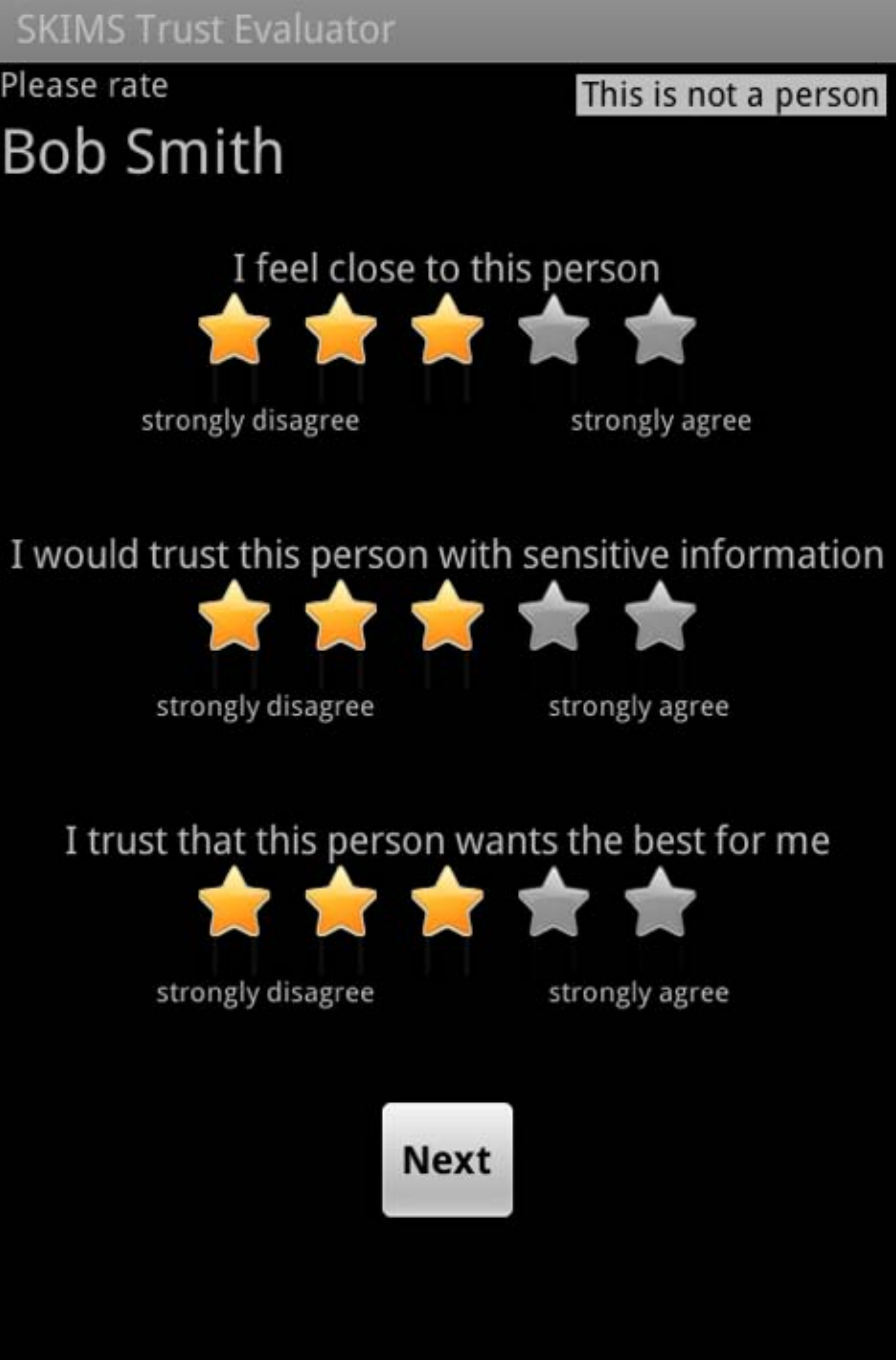}\label{img:ratingScreen}}
  \quad
  \subfigure[Sample result in YAML format]{\includegraphics[width=0.45\columnwidth]{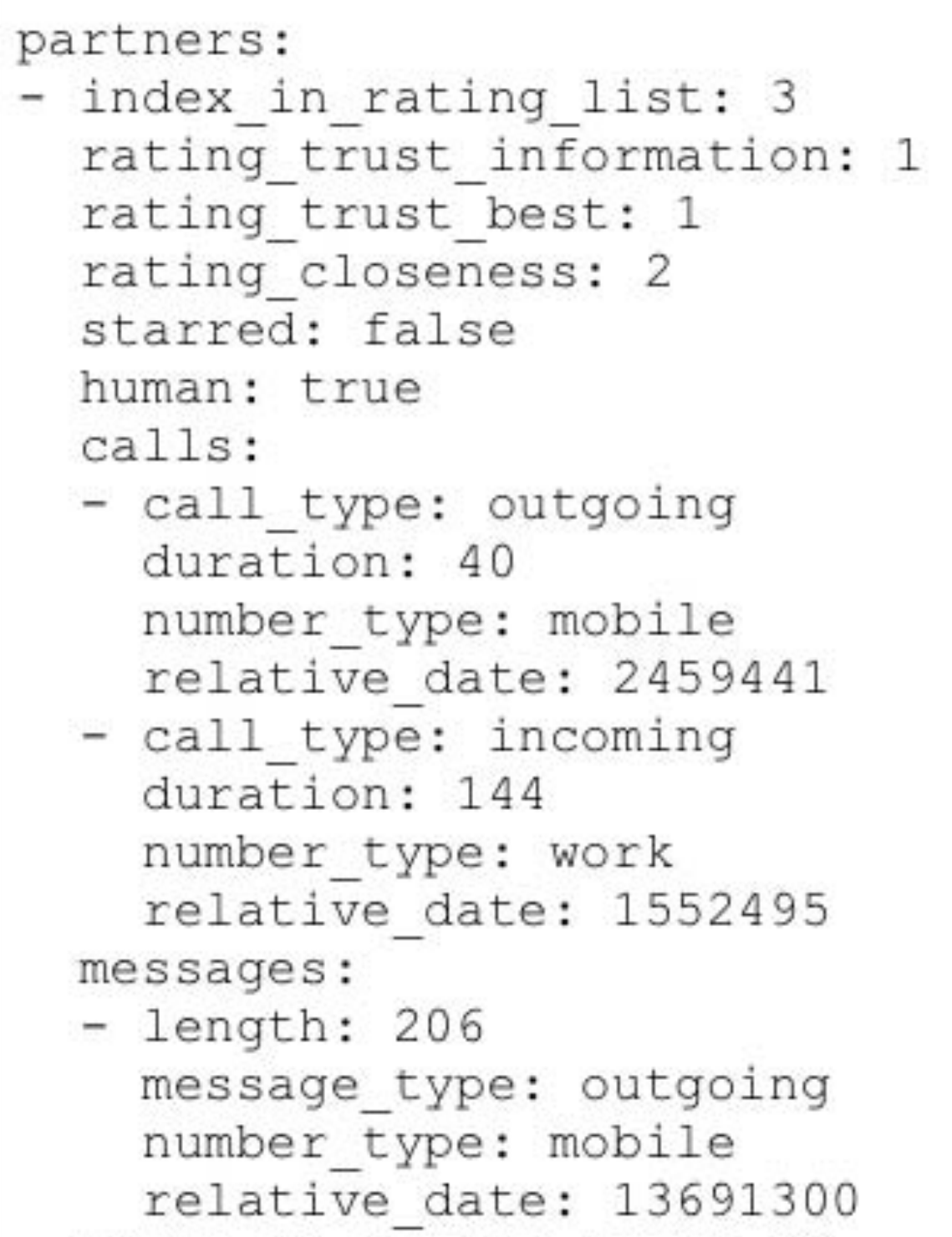}\label{img:yamlResults}}
  \caption{Survey app to rate contact list entries }\label{img:surveyApp}
\end{figure}

The survey application first shows a welcome screen while analyzing the
call and message logs of the phone at the same time. All contacts that
appear in those logs are denoted \emph{(communication) partners} and are collected and ranked based on the number of
interactions with them (i.e., combined number of calls and messages). On
the next screens (cf., Figure~\ref{img:ratingScreen}) the participant will
be asked to rate 20 different partners. The partners prompted are
alternately assigned from three classes: contacts that (a) have the
\emph{most interactions} with the participant, (b) have the \emph{least
interactions} (but not zero), and (c) are \emph{randomly} selected from the
list of partners. For every contact, the subject should give a response to
each of the three closeness/trust statements. Limiting the number of
contacts to rank is important to not overload participants with questions. Our
measurements show that some people have more than 160 communication partners. In
addition, a pure random selection on a limited list may lead to convergence
problems, which we avoid with our selection strategy.

To exclude common business contacts in the survey, participants have been
asked to use the button labeled \emph{This is not a person}, which skips
the current contact. After all 20 contacts are rated the phone's email
client presents the results in a message that will be sent.

Respecting the participants' privacy and acting transparently was a huge
concern when developing the app. As a consequence the results have been
made anonymous. Transparency has been implemented by sending results via
email and not by an upload in the background that can not be observed and controlled by the user. The participants can decide
whether or not to send the email. Furthermore, the results have been
encoded in YAML (cf., Figure~\ref{img:yamlResults}), a format easily
readable by humans. A beta phase revealed that XML was hard to understand
by the subjects. In addition, Yahoo's email client on Android removes all
XML tags from the email leaving it unusable for further processing.

\subsubsection{Acquired Data}

The results of the questionnaire consist of two parts: (a) general statistics and (b) details about
all communication partners. The \emph{general section} includes the number
of entries in the address book, the amount of calls and messages in the
corresponding logs, and the number of active communication partners. The
\emph{list of communication partners} contains meta data for every entry (i.e.,
if applicable, position in the questionnaire and rating; human state;
favorite state in the address book), as well as a list of all calls and
messages  recorded in the logs.  For every call to or from this
communication partner the relative date, its type (i.e., incoming,
outgoing, or missed), the duration in seconds, and potential personal tags (e.g.,
\texttt{home} or \texttt{work}) are stored (cf., Figure~\ref{img:yamlResults}).  The date is concealed for
privacy reasons. It represents the number of seconds between the time of
the call and the time the survey app is executed extended by a random
offset that is locally created for every participant. Analogously the
listing of messages contains the message type, the tag assigned  to the number, the
relative date, and the length of the message in characters.

\subsubsection{Recruiting via Amazon Mechanical Turk}

To reach a greater number of participants and to avoid a bias by asking
only personal acquaintances, we published the survey on the crowd-sourcing
platform Amazon Mechanical Turk (AMT). The concept behind it is simple: A
requester offers a Human Intelligence Task (HIT) that will be claimed by
globally distributed workers, which get a compensation for a successful
HIT. A single HIT can be handled by multiple people. AMT is
well-established in the scientific community, in particular in the context
of social science~\cite{b-ssp-11}, but it gets popular as an appropriate
tool for  computer scientist, as well \cite{lgkm-afpsu-11}. A major concern of such an approach
might be the trust in the reliability of the (paid) answers.  However,
recent studies showed that survey results by AMT workers are consistent in
quality compared to classic studies~\cite{b-ssp-11}.

Our survey requires workers to download and install the Android app,
perform the survey, and send the results in an email combined with their
worker ID, so the task could be approved via the AMT platform. We calculate
\$1 per worker, which is slightly above-average but participants
need to install the app. It is worth noting, that an overpriced task may
reduce the result quality~\cite{mw-fipc-09}.

\begin{figure*}[ht]
  \center
  \subfigure[Calling frequency]{\includegraphics[width=0.4\columnwidth]{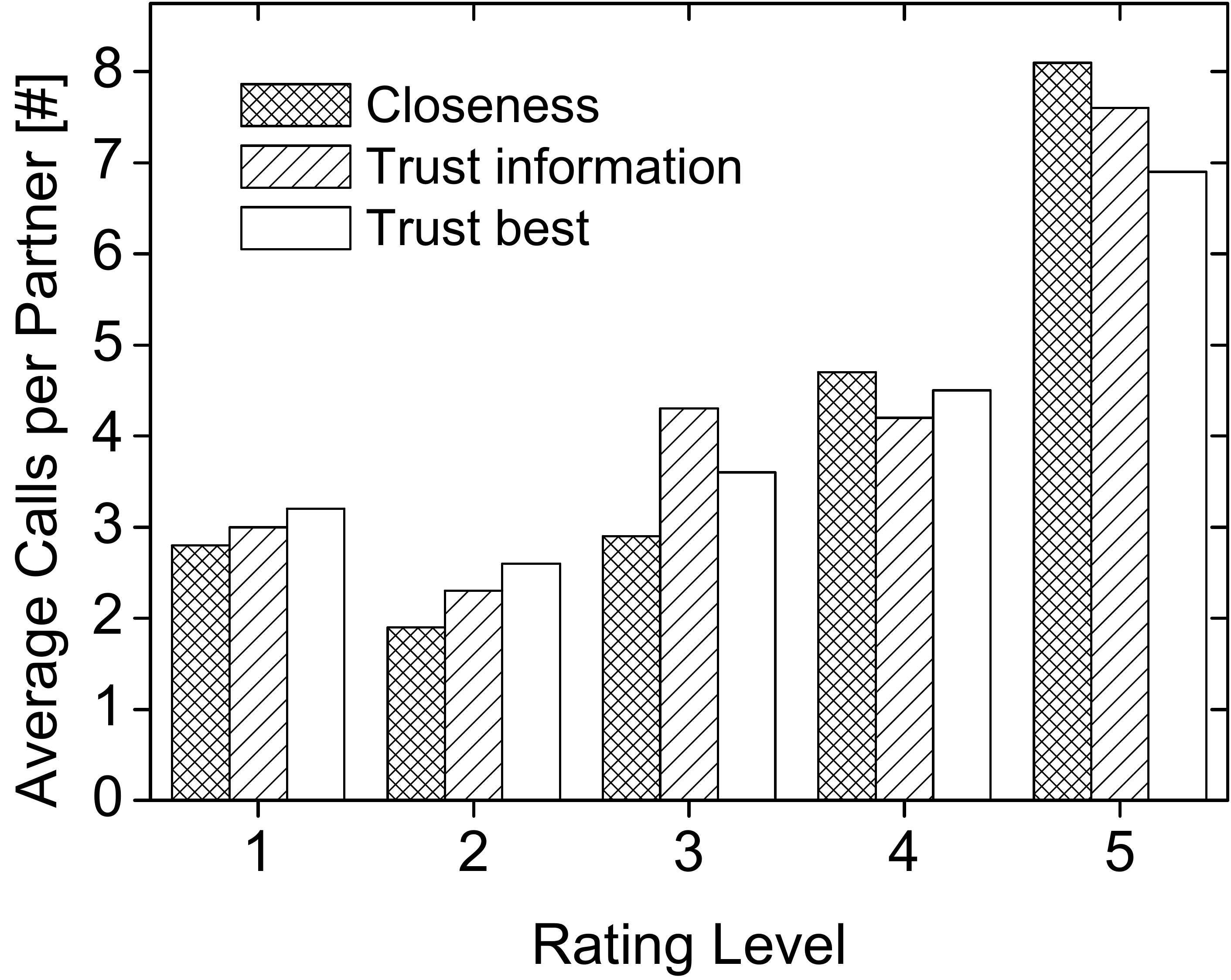}}
  \qquad
  \subfigure[Messaging frequency]{\includegraphics[width=0.4\columnwidth]{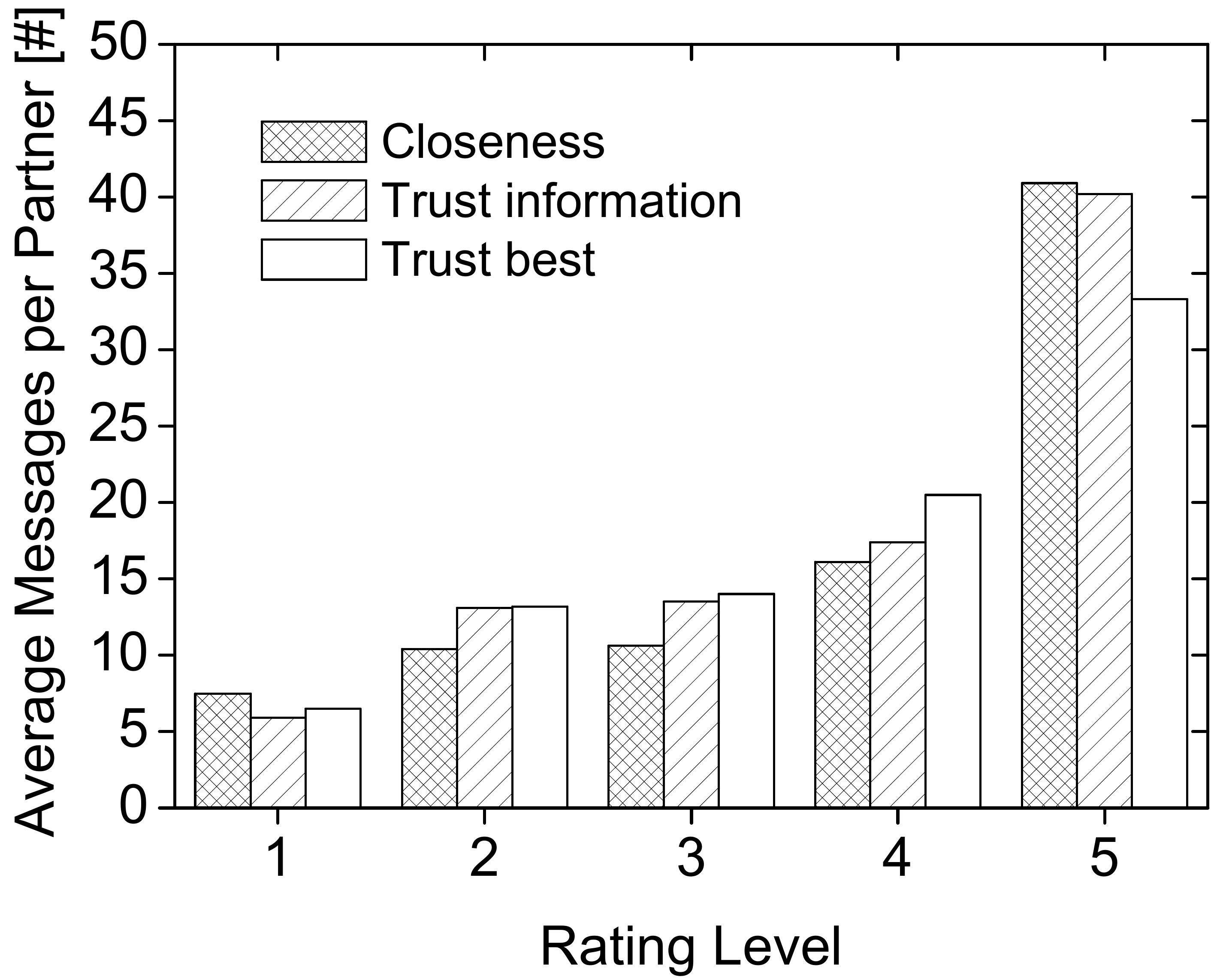}}
  \qquad
  \subfigure[Call duration]{\includegraphics[width=0.4\columnwidth]{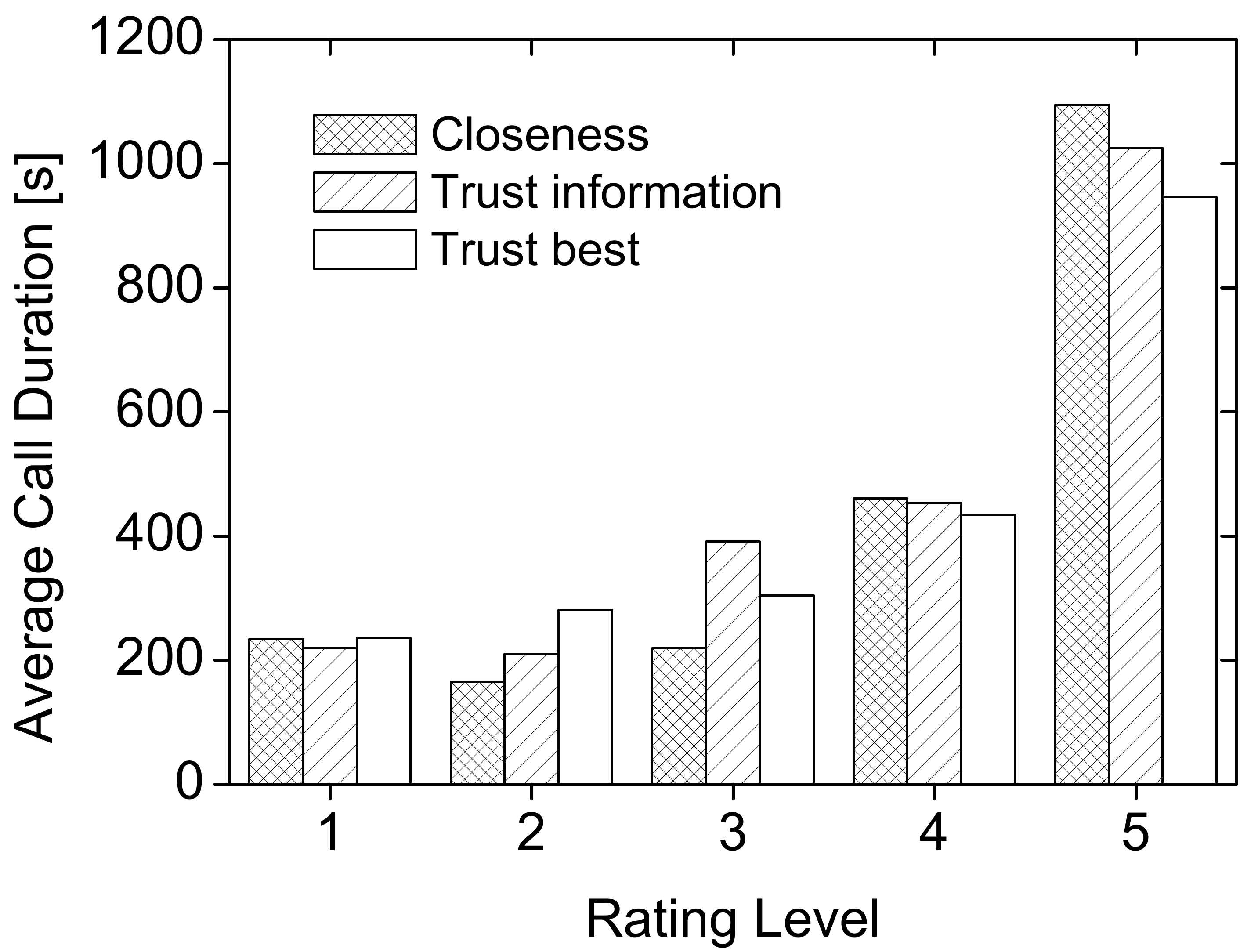}}
  \qquad
  \subfigure[Message length]{\includegraphics[width=0.4\columnwidth]{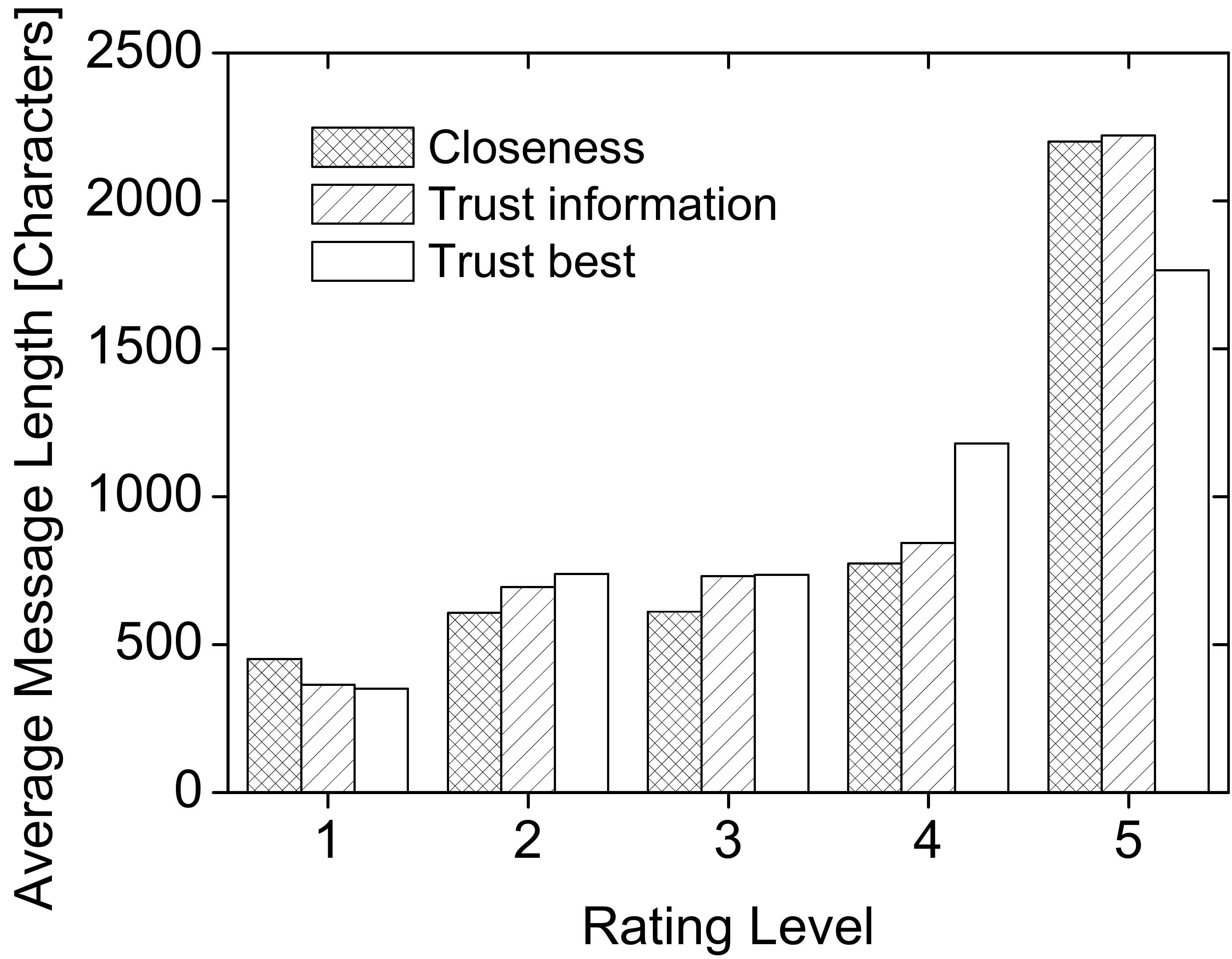}}
   \caption{Average amount of communication depending on the rating of
   trust and closeness (1 = complete disagree, \dots, 5 = strong agree)}\label{img:absoluteValues}
\end{figure*}

\subsection{Verifying Quality of Answers}  

We started our survey in December 2011. In total, we received \numResultsAMT results
after 6 weeks. Some workers completed the survey more than once, especially
when they forgot to transmit their worker ID in the first email. This led
to 37 ``double'' submissions, which have not been considered in the
analysis to avoid statistical weights for dedicated participants. Overall
our subsequent analysis is based on \numResultsAMTUnique answers.

On the AMT webpage where workers claim the HIT, we allowed participants to
share their opinions and thoughts. Surprisingly we received a lot of
feedback from several users, which helped us to improve the survey and to
verify the seriousness of the answers. 
At the beginning, a typical feedback was such as \emph{``There should be an
option to discart some of the contacts. For example, the app asked me about
a pizza place that i had in contact list\dots it doesn't make sense to
answer the questions in this case.''}. We already provided such a
function but  thereupon increased its visibility. We then asked the concerned
participants to repeat the survey, which all of them did. In the end,
we could not identify a significant difference in the number of
non-personal contacts before and after we applied the small change in the application. This indicates the statistical
insignificance of those events.

Overall, the feedback of the participants gives us confidence that they
took the survey seriously and did not randomly complete the form. Amongst
others, this behaviour is motivated by a higher reputation for future
HITs~\cite{mw-fipc-09}. On the other hand, three out of \numResultsAMTUnique participants
(1.3\%) skipped through the contacts without giving a single rating.
Those results were excluded from further investigation.

As a site observation we noted that participants have only very limited
privacy concerns. Only positive comments on this topic (\emph{``ya survey was very nice and did not hurt me in any way.''}) have been raised and only a
very few downloads of the source code have been initiated.

\subsection{General Observations} \label{ssec:generalObservations}

After excluding all double and invalid submissions, we consider the data of
\numResultsAMTHonest participants, which overall includes information about \numPartnersAMT
communication partners that performed \numCallsAMT calls and wrote and received
\numMsgsAMT messages. For \numPartnersRatedAMT contacts (48,4\%) we received personal ratings
of trust and closeness.
The size of the contact list ranges per participant from 2 up
to 1,827 entries with a mean value of 249.3. The standard deviation
of 308.7 indicates high fluctuation. However, the
address book may include outdated entries. On average 32 \emph{different}
communication partners appear in the call and message logs that show active
parties, and an average of 304.6 calls and 739.2 messages have been handled.

A requirement for this study is the collection of a representative
observation period to reflect common mobile usage. Datasets that include
only logs spanning short time periods may skew the results in particular if
they contain very limited communication partners. The typical interaction
log covers on average 3 months of activities. Similar to the contact list
sizes, the sample intervals are widely spread with a standard deviation up
to 135.9 days. Around 88\% of the logs provide data for more than one week
and 70\% reveal interaction for more than one month. Six participants
exhibit very short-dated logs ($\le$1~day). After asking the owners we can
ensure that this is either coincidentally due to a reset of the mobile
shortly before the survey started or a recent purchase. Usually, these
cases exhibit only few communication partners. However, some logs show a
high amount of activity during a short period. One result, for example,
documents 23 different communication partners in less than 24 hours. To
minimize skewed results, we decided to exclude logs that last less 
than one week and contain less than 10 communication partners as well as those logs that contain no more than 4 different communication partners.
The results of the 29 affected participants were excluded from further analysis. 

For the analysis performed in the following sections we take as a basis the survey results cleared from unreliable or double submissions. As described above we also filtered out results that showed typical patterns for phones that were recently reset or put into operation. Thus, in the following sections we work with the reliable and significant results of \numResultsAMTClean participants providing information about \numCallsAMTClean calls and \numMsgsAMTClean messages from \numPartnersAMTClean communication partners of which \numPartnersRatedAMTClean have been rated.

\section{Survey Evaluation} \label{sec:evaluation}

The results obtained from the Android survey with \numResultsAMTClean significant data sets can be evaluated in different ways. 
To answer the question if and how it is possible to predict closeness or
trust based on available activity logs we identify appropriate indicators using
absolute as well as relative measurements. We further incorporate manual
tags by the mobile phone's user.

\subsection{Absolute indicators} \label{ssec:absoluteValues}

In general, the level of trust and closeness is independent of the type of
communication (voice vs. text) but depends significantly on the amount of
communication (cf., Figure~\ref{img:absoluteValues}). The top most
communication partners are ranked with the highest trust value. On average
$\approx$~8~calls and $\approx$~40~messages have been exchanged by our
surveyed users with very trustworthy or close parties. 

To analyze the relationship between trust and closeness in more detail,
first, we calculate pairwise correlations. As shown in
Table~\ref{tab:correlations}, all correlations are high and statistically
significant.
\begin{table}[hb]
\centering
\small\begin {tabular}{ c c  | c } 
 \hline
\textit{RatingCloseness}&\textit{RatingTrustInfo}&0.7976\\
\textit{RatingCloseness}&\textit{RatingTrustBest}&0.7465\\
\textit{RatingTrustInfo}&\textit{RatingTrustBest}&0.7834\\
\hline
\multicolumn{3}{l}{\footnotesize{All correlations are statistically significant (99\%)}}\\ 
	\end {tabular}
	\caption{Correlations between rating variables}
\label{tab:correlations}
\end {table}	

\begin{figure}[b]
  \center
  \includegraphics[width=0.8\columnwidth]{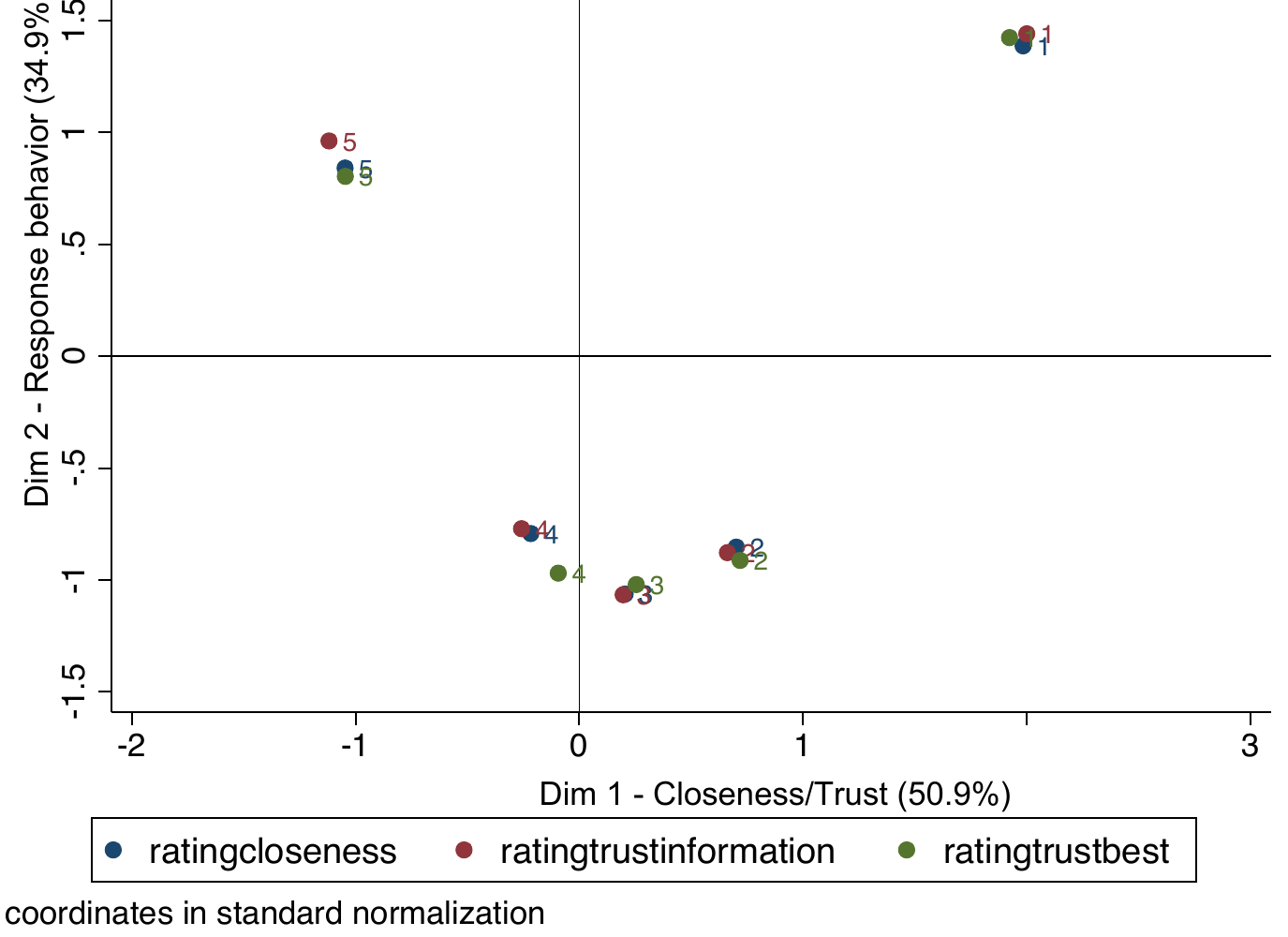}
   \caption{Multiple correspondence analysis of the rating variables}\label{img:ratingCorrelation}
\end{figure}

As correlations only have limited validity in the analysis of categorical
data beyond that, we also conduct a multiple correspondence analysis (MCA). 
In a nutshell, MCA is an exploratory technique which serves to find common
dimensions within a group of variables. Figure~\ref{img:ratingCorrelation}
shows the results for our three survey questions. 
The first dimension reflects the closeness/trust-dimension, with categories
1 to 5 ordered along the x-axis. Note that categories 1 and 5 are further away from categories 2 to 4, which lie relatively closer to each other. The second dimension reflects the response behavior of the participants, ranging from moderate to extreme. Such a dimension is common among data collected by means of a Likert scale. 
Most importantly, the results show high similarity between each of the
categories of the three questions, which, together with the high
correlations between the variables indicates that closeness and trust
towards a person are highly interchangeably. Thus, tie strength and
closeness (cf., Section~\ref{sec:related-work}) can be transferred into
trust regarding sensitive information and overall goodwill. Throughout our
subsequent analysis, we will only show answers to the question \emph{I
would trust this person with sensitive information} as replies to the other
two questions exhibit similar values.

\begin{figure}
  \center
  \includegraphics[width=0.7\columnwidth]{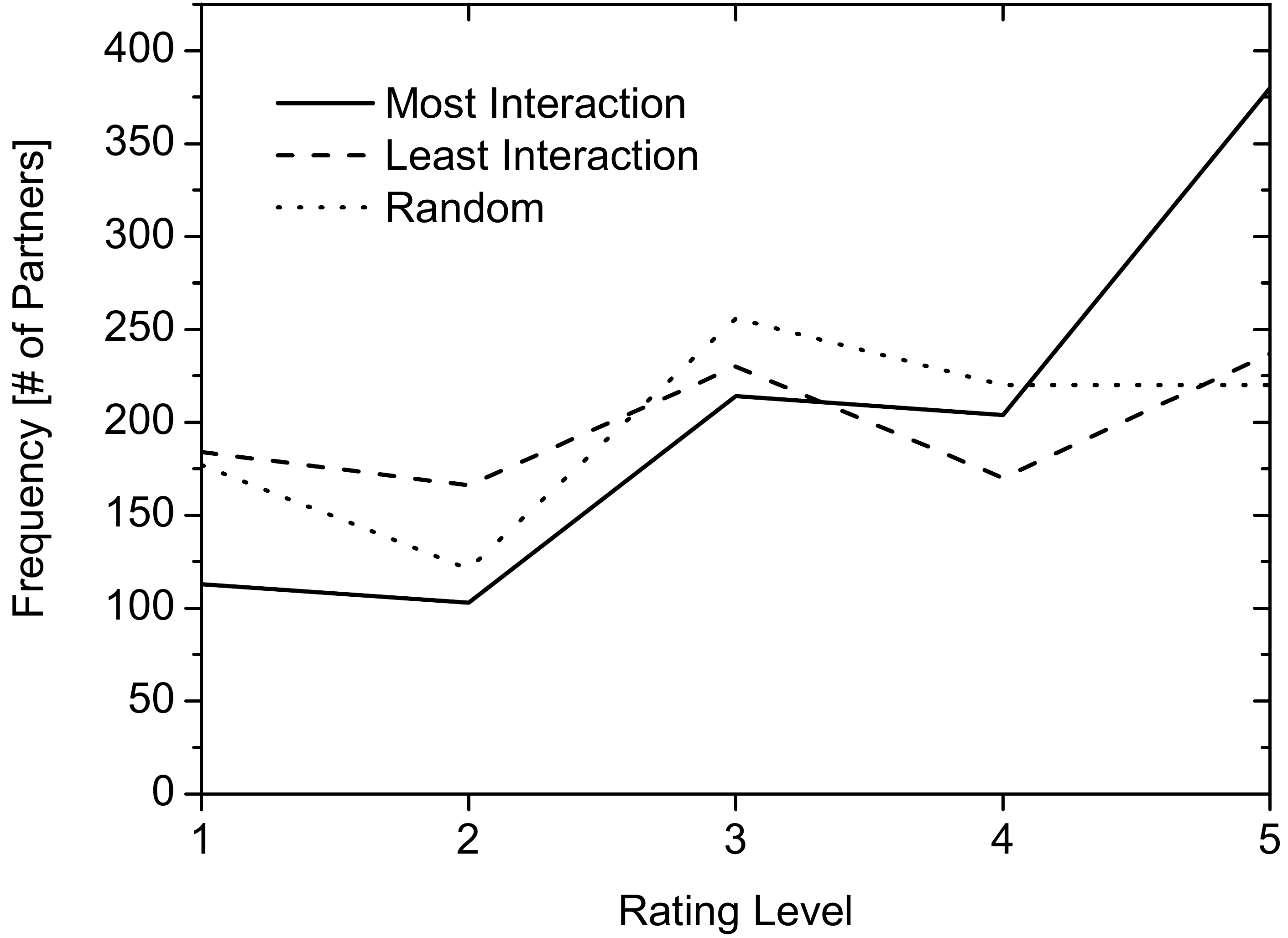}
   \caption{Distribution of ratings for three different interaction classes
   over all survey participants}\label{img:classes}
\end{figure}

Accumulated over all surveyed users, Figure~\ref{img:classes} shows the
frequency distribution of the given ratings for the three interaction
classes \emph{most interactions}, \emph{least interactions}, and
\emph{random} (cf., Section~\ref{ssec:application}). The number of
communication partners with the least interactions and randomly selected
peers are more evenly distributed across the rating levels.  Partners that
belong to the high interaction class exhibit an increased probability for
a strong agreement on trust, in which lower rating levels are less
pronounced.  A high number of interactions, thus, may indicate higher trust
whereas less interacting peers cannot be categorized as trustworthy.
Note that significantly less partners have been rated with a trust level of
2 than all other trust levels. This illustrates that the subjective level
of non-negative trust towards a person (level 4 or 5) allows for improved
differentiation compared to the level distrust (level 1 or~2).

\subsection{Manual Tags}

Smartphones provide users with the option to manually add tags to their
contact lists. Contacts can be marked as favorites to appear in a special
list or annotated with dedicated labels (e.g., \emph{home}, \emph{work}, or
\emph{mobile}) to group and ease lookup. Our analysis show that those
labels are not helpful for trust prediction, since they are not widely used (only 36.6\% of the participants tagged at least one number in their contact list) and if a number is tagged its mostly labeled with \emph{mobile} (76.2\%).  The favorite list, however, is a good indicator for trust.
Out of \numResultsAMTClean participants 97 selected at least one
contact as favorite. On average, every participant that applies this feature
tags $\approx$~4 communication partners. Table~\ref{tab:favorites}
shows the probability that a favorite contact attains a specific rating
level. A favorite contact enjoys a trust value of 3 or
higher with almost 90\% probability. This feature allows for the
identification of trustworthy partners with high confidence.

\begin{table}[]
\centering
\small\begin {tabular}{ c | c | c } 
Rating Level&Percent&Cum. Percent\\
\hline
5 & 60.53 & 60.53 \\
4& 19.55 & 80.08  \\
3 & 9.02 & 89.10 \\
2 & 5.64 & 94.74 \\
1 & 5.26 & 100.00 \\
\hline
\end {tabular}
\caption{Rating levels for contacts tagged as \emph{favorites}}
\label{tab:favorites}
\end {table}	

\subsection{Relative indicators} \label{ssec:relativeValues}

\begin{figure}
  \center
  \includegraphics[width=0.7\columnwidth]{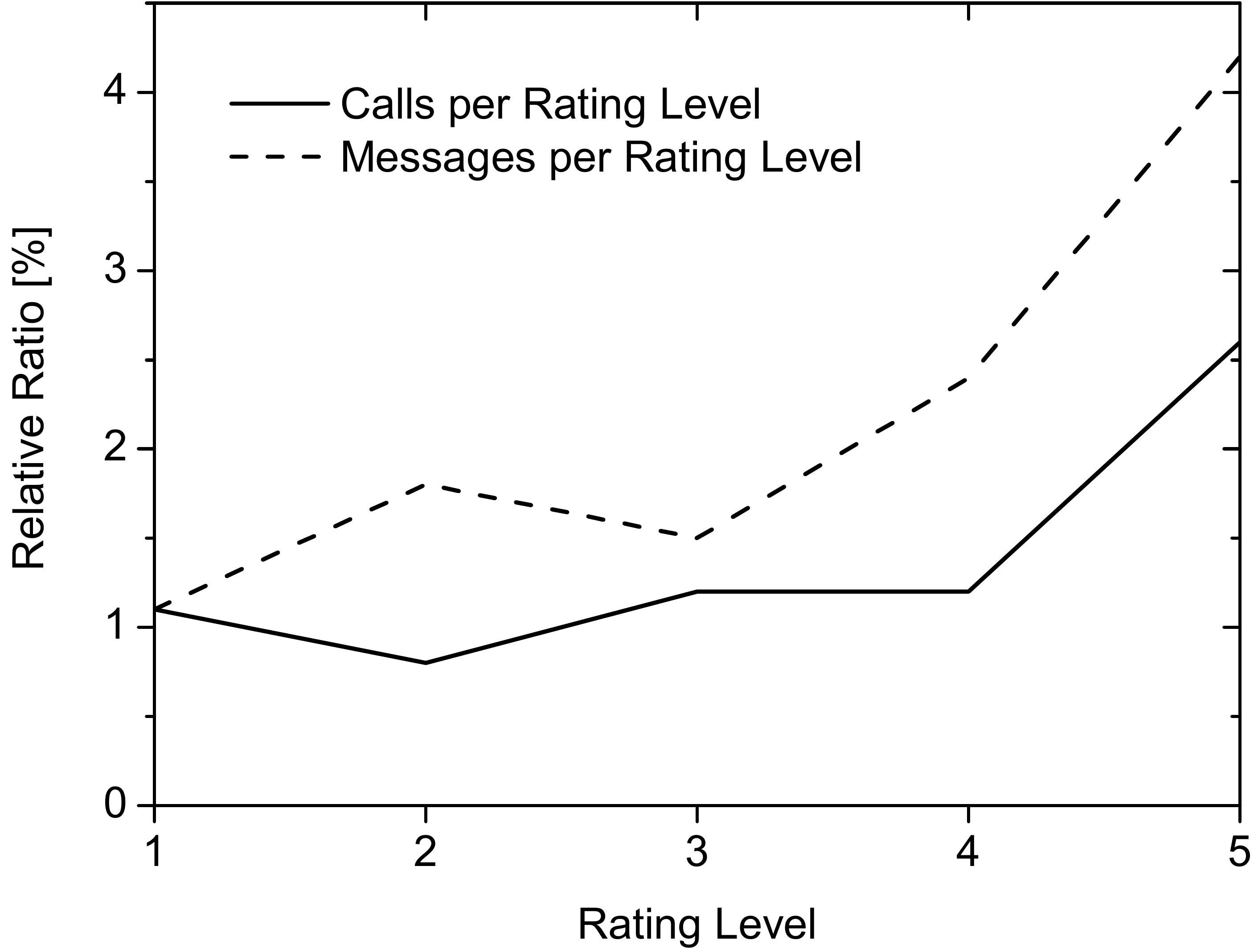}
   \caption{Proportion of the total interactions per participant}\label{img:relativeValues}
\end{figure}

Besides the absolute values examined in Section~\ref{ssec:absoluteValues}, relative variables can be good indicators for trust as well. Useful insights can be gained by examining the proportion of the number of calls (or messages) per partner to the total number of calls (or messages) of a survey participant. Once again the lower rating levels do not show a trend (cf., Figure~\ref{img:relativeValues}). 
The fraction of messages of the total number of messages increases with the trust levels 4 and 5. For the fraction of calls this is only true for level 5.

Values that do not prove to be good indicators include the average call duration and the average message length (cf., Section~\ref{ssec:absoluteValues}). Both do not show the expected trend, that means more trust does not result in longer average calls. This might be due to the fact that especially the number of calls to trusted individuals is high, thus leveraging the mean average call duration.
Investigating temporally relative values, e.g., number of calls per hour, neither provides new findings. The number of calls and messages do not correlate with the duration of the log, so longer logs do not generally mean more interaction. Still, the number of interactions per time neither serves as a predictor for trust.

\section{Application Perspective} \label{sec:application}

The results of our survey can be helpful in providing new ways of looking at the socio-technical interface especially in mobile phone MANETs.

\subsection{Trust Establishment for Smartphones}

In previous work \cite{tws-spypt-11} it was shown that trust
establishment between two mobile phones can be based solely on social data inherently available on those devices. When two phones need to establish a trust relationship they do so on basis of the mutual contacts in their contact lists. Finding mutual contacts is a variety of the Private Set Intersection Problem which has been subject to many publications. As soon as the mutual contact list entries are found each of the two mobile phones individually examines the results. The plain amount of those mutual contacts can give a rough indication of the relationship between the two phones' users. More insights can be gained with a qualitative analysis. 

When two phones try to establish a trust relationship amongst each other they can base this decision on the combined trust towards the contacts they both share. Therefore, each phone calculates a trust value to each of the mutual contacts. To be functional even without central infrastructure the phone can only rely on data inherently available on mobile phones. Consequently, call and message logs serve as a foundation for trust evaluation. The combination of the trust values towards each mutual contact results in a trust estimation of the other device. Note, that both devices perform an individual trust estimation that may result in asymmetrical trust estimates.

\subsection{Finding a metric} \label{ssec:findingMetric}

The analysis show that several metrics can be derived from a mobile phone's call and message logs that can indicate a trust relationship to  specific communication partners. Trust prediction is fairly easy were high absolute (i.e., number of calls and messages to and from the partner, cumulated call duration and message length) or relative (i.e., fraction of calls and messages of the user's total) values can be observed.
The data is not that evident for lower numbers and lower trust levels. Whether a survey participant rated a communication partner with a level of 2 or 3 seems to be a very subjective matter.

\begin{table}[b]
\centering
\small\begin {tabular}{ r | c | c | c | c } 
 &\multicolumn{3}{|c}{Quantiles}\\
Variable name&75\%&90\%&95\%&99\%\\
\hline
\texttt{Number of calls}&2.0&7.0&13.0&42.0\\
\texttt{Number of messages}&2.5&14.0&35.0&84.0\\
\texttt{Duration of calls}&89&460&711&4759\\
\texttt{Message length }&158&1003&2105&5580\\

\texttt{Rel number of calls}&0.6&2.6&5.5&14.6\\
\texttt{Rel number of msgs}&0.6&2.3&5.9&17.4\\
\hline
\end {tabular}
\caption{Quantiles of selected variables for rating level 1}
\label{tab:quantiles}
\end {table}	

A metric to predict trust should take into account all of the parameters mentioned. We create a metric by looking at the quantiles for each of the parameters where the trust level is 1 (cf., Table \ref{tab:quantiles}). Taking, as an example, the 95\% quantile and \texttt{number of calls} into account, this means that 95\% of all partners with a rating level of 1 have had less than 13 calls with the user. 
Forming a condition joined by the OR-operator combining all variables, we can gradually assign trust by comparing the different quantiles. If the values of the 95\% quantiles are chosen as thresholds, more than 80\% of the communication partners, that satisfy at least one of the conditions, have a rating level of 3 or more (cf., Table \ref{tab:metric}).
Using higher (or lower) quantiles as thresholds will increase (or decrease) this value.
Even more accuracy can be gained if the conditions are combined pairwise with an AND-operator. For example, if $\texttt{Number of calls} > 13$ AND $ \texttt{Message length} > 2105$ more than 85\% of the resulting communication partners have a rating level of 3 or higher. Excluding all partners that were not tagged as a favorite by the user 100\% of the results have a trust level of 3 or higher.

\begin{table}[ht]
\centering
\small\begin {tabular}{ c | c | c } 

Rating Level&Percent&Cum. Percent\\
\hline
5 & 42.33 & 42.33 \\
4&19.33 & 61.66  \\
3 & 18.90 & 80.56 \\
2 & 8.86 & 89.42 \\
1 & 10.58 & 100.00 \\
\hline
\end {tabular}
\caption{Distribution of rating levels for a joined OR condition with thresholds of the 95\% quantiles}
\label{tab:metric}
\end {table}	

Concluding it can be seen that trust can be predicted with data typically available on every mobile phone. Merely new or recently reset mobile phones might not be able to provide a minimum data foundation. As shown in the survey the mean log duration is at about 110 days, so we are confident to find enough useable data in general.

\subsection{Comparison with existing work}

Social data available on mobile phones can be used in order to predict
future callers and callees \cite{pd-tucwc-11}. Analogously to the trust
establishment approach in Section~\ref{ssec:findingMetric}, the authors define a metric to rate the closeness between user and contact. 
The authors classify all contacts in one of three groups of different closeness by an algorithm based on the call activity with each contact.
Applying this metric to our data set can verify the algorithm or find weaknesses. 

The call prediction metric associates the contacts into three groups ``Socially Closest'', ``Socially Near'', and ``Socially Distant''. In order to perform a meaningful comparison with our survey a mapping to the five point scale we used is needed. The character of the mapping obviously influences the outcome of the comparison. 
As a first, intuitive mapping a rating of 1 or 2 was assigned to the ``Distant'' group, a rating of 3 was mapped to the ``Near'', and a rating of 4 or 5 represented the ``Closest'' communication partners. Then for every rated communication partner this binning was compared to the outcome of the metric. With this binning assumption the mean error for all survey participants comes to $0.86$ meaning that 86\% of the communication partners were evaluated in a lower group than the survey results support. Since the authors only used call activity as a basis for their metric \cite{pd-tucwc-11} the figures above only represent this original algorithm. Extending the metric to include messaging activity as well lowers the mean error to $0.77$. Choosing a different binning also reduces the mean error. When mapping a rating of 5 to the ``closest'' class, a rating of 4 to the ``near'' class, and a rating of 3 or less to the ``distant'' class the mean error can be lowered to $0.46$.  Taking also into account the messaging activity this can even further be lowered to $0.38$, meaning that about 38\% of the contacts in a contact list were assigned a closeness evaluation of a lower group. Thus, the analyzed metric tends to underestimate the closeness of a user to its contacts. Furthermore, using only call data and not considering messaging activity also reduces its accuracy. This might also be due to the fact that the authors only surveyed 10 individuals in regard to their closeness towards contacts. With the data collected in the scope of this work the metric can be improved.

\section{Conclusion \& Outlook}\label{sec:conclusion}

In this paper we presented data from a large-scale survey of Android users that helps to bridge the gap between data that can be found on mobile phones and social relationships users of mobiles share. By analyzing more than 200 contact lists and interaction logs 
we identified metrics that can be used to evaluate trust towards peers from a user's contact list using absolute as well as relative indicators. 
The results of the survey also indicated that closeness and trust are highly correlated, thus linking previous research with our findings.

The findings can help to improve application scenarios like trust establishment in mobile ad-hoc networks of mobile phones, since they inherit the social data needed.

\clearpage

\bibliographystyle{acm}
\bibliography{/bib/own,/bib/rfcs,/bib/ids,/bib/cloud,/bib/theory,/bib/layer2,/bib/internet,/bib/transport,/bib/overlay,/bib/vcoip,/bib/visualization,/bib/security,/bib/manet,/bib/socio}


\end{document}